\documentstyle[pra,aps,multicol,psfig]{revtex}

\def\addcontentsline#1#2#3{\relax}
\begin{document}
\draft
\title{Impurity pinning in transport through  
1D Mott-Hubbard and spin gap insulators. }
\author{V.V. Ponomarenko$^*$ and N. Nagaosa}
\address{Department of Applied Physics, University of Tokyo,
Bunkyo-ku, Tokyo 113, Japan}
\date{\today}
\maketitle
\begin{abstract}
A low energy crossover \cite{1} induced by Fermi liquid reservoirs 
in transport through a 1D Mott-Hubbard insulator of finite length $L$ 
is examined  in the presence of impurity pinning. Under the assumption that  
the Hubbard gap $2M$ is large enough: $M > T_L \equiv v_c/L$ 
($v_c$: charge velocity in the wire) and 
the impurity backscattering rate $\Gamma_1 \ll T_L$,
the conductance vs. voltage/temperature displays a zero-energy
resonance. 
Transport through a spin gapped 1D system is also described 
availing of duality between 
the backscattered current of this system and 
the direct current of the Mott-Hubbard insulator. 

\end{abstract}


%
\multicols{2}
Recent developments in the nano-fabrication technique have allowed   
to produce relatively clean one channel wires of 
$1-10 \mu m$ length. In these wires, 1D Tomonaga-Luttinger Liquid (TLL)
behavior \cite{hald} was observed in transport measurements \cite{tar,ya}. 
Then it has been suggested that the correlated insulating behavior may be
tuned with this setup \cite{ftr}. 
This behavior of the 1D electron systems is 
expected \cite{sol} at half-filling where any Umklapp scattering has to
open a Hubbard gap $2M$ in the charge mode spectrum of the infinite wire
if the forward scattering is repulsive.  
Earlier consideration  \cite{1} predicted  
crossover from Fermi liquid transport carried by the 
reservoirs electrons at low energy to that of a Mott-Hubbard insulator in 
the one channel wire of long enough length $L$:
$M > T_L \equiv v_c/L$ ($v_c$: charge velocity in the wire).
Charge in this insulator is carried by 
soliton excitations of the condensate emerging in the charge mode. 
This insulator, which is probably the
easiest for the experimental observation,
features two marginal quantities  \cite{2}: charge of the soliton is 
unchanged $e$ ($e=\hbar=1$ below) and the exponent $1/2$ characterizing
the electron-soliton transition brings about similarity with the
free electron tunneling through a resonant level. 
  
In this paper we investigate the effect of impurity
backscattering of low rate $\Gamma_1 \ll M$ on the above crossover. 
For low energy following Schmid \cite{schmid} the problem is mapped through
a Duality Transform onto the model of 
a point scatterer with pseudospin imbedded in TLL. It is solved
exactly by fermionization. 
The result shows that the impurity enlarges the width of a zero energy resonance in the conductance up to $\Gamma_1+\Gamma_2$ 
where the exponentially small $\Gamma_2$ ($ \Gamma_2 \approx \sqrt{T_L M}
e^{-2M /T_L}$ at half filling) is the rate of tunneling of the condensate phase. 
The linear bias conductance at zero temperature 
$G={\Gamma_2 \over \pi (\Gamma_1+\Gamma_2)}$ reveals an exponential  
enhancement of the amplitude of the impurity potential
$\sqrt{ \Gamma_1 / \Gamma_2 }$ by the condensate tunneling. 
This suppression of 
the resonant conductance does not affect the saturation of the
current at $J=\Gamma_2 $ above the crossover as far as
$\Gamma_1+\Gamma_2 \ll T_L$.
Finally, the result is addressed to the case when the gap opens
in the spin mode of the wire. It occurs if the effective interaction
between electrons inside the wire is attractive \cite{sol}.
Although it is not expected for the semiconductor nanostructure, it 
has been observed in quasi 1D organics \cite{jer}. The 
current backscattering in this system coincides with the direct one 
of the Mott-Hubbard insulator multiplied
by a factor ${ \Gamma'_1 \over \Gamma'_2}$, where $\Gamma'_i$ has the same
meaning for this system as $\Gamma_i$ above.
So, the  backscattering
current is zero in the absence of impurities, and 
the current is maximum $V/\pi$. 
Impurity appearance gives rise to backscattering together with 
simultaneous tunneling between the spin mode vacua of the wire. 
The conductance
is suppressed below a crossover energy and recovers above it. It
results in a constant shift $\Delta J=-\Gamma'_1$ of the current at
high voltage. 

Transport through a 1D channel wire confined between
two leads could be modeled by a 1D system of electrons
whose pairwise interaction is local and switched off outside the finite
length of the wire. Applying bosonization and spin-charge separation
we can describe the charge and spin density fluctuations
$\rho_{b}(x,t)=(\partial_x \phi_{b}(x,t))/(\sqrt{2} \pi), \ b=c,s$,
respectively, with (charge and spin) bosonic fields $\phi_{c,s}$. 
Without impurities
their Lagrangian symmetrical under the spin rotation reads
\endmulticols
\vspace{-6mm}\noindent\underline{\hspace{87mm}}
\begin{equation}
{\cal L}= \int dx \sum_{b=c,s}
\bigl[ \frac{v_b(x)}{2g_b(x)}
\{ 
{1 \over v_b^2} \left({{\partial_t \phi_b(t,x) }
     \over {\sqrt{4 \pi}}}
\right)^2 -
\left({{\partial_x \phi_b(t,x) } \over {\sqrt{4 \pi}}} \right)^2 
\}
-
{E_F^2 U_b \over \pi v_F} \varphi(x) \cos({2 \mu_b \over v_b} x +
\sqrt{2} \phi_c(t,x)) \bigr]
\label{1}
\end{equation}
\noindent\hspace{92mm}\underline{\hspace{87mm}}\vspace{-3mm}
\multicols{2}
\noindent
where $\varphi(x)= \theta (x) \theta (L-x)$ specifies a one channel wire
of the length $L$ adiabatically attached to the leads $x>L,x<0$,
and $v_F(E_F)$ denotes the Fermi velocity(energy) in the channel. 
The parameter $\mu_c \equiv \mu$ varies the chemical potential inside
the wire from its zero value at half-filling and $\mu_s=0$. 
The constants of the forward scattering differ inside the wire 
$g_b(x)=g_b$ for $x \in [0,L]$ from those in the leads 
$g_b(x) =1$, and an Umklapp scattering (backscattering) of 
the strength $U_c (U_s)$ is introduced inside the wire. The velocities
$v_{c,s}(x)$ change from $v_F$ outside the wire to some constants $v_{c,s}$
inside it. We can eliminate them rescaling the spacial coordinate 
$x_{old}$ in the charge and spin Lagrangians of (\ref{1}) into 
$x_{new}\equiv \int^{x_{old}}_0 dy/v_{c,s}(y)$.
As a result, the new coordinate will have an inverse energy dimension 
and the length of the wire becomes different for the charge mode $L \to 1/T_L$ 
and spin mode $L \to 1/T_L'$. 
Applying renormalization-group results of the 
uniform sin-Gordon model at energies larger than $T_L$ or $T_L'$ 
we come to the renormalized values of the parameters in (\ref{1}). 
For the repulsive interaction when initially
$g_s>1>g_c$, the constant $U_s$ of backscattering flows 
to zero and $g_s$ to 1, bringing the spin mode into the regime of the
free TLL . The constant $U_c$ of Umklapp process
increases, reaching $v_F/v_c$ at
the energy cut-off corresponding to the mass of the soliton $M$ if the 
chemical potential $\mu$ is less than $M$. Meanwhile, $g_c$ flows to its free 
fermion value $g_c=1/2$ \cite{1}. These values of the parameters specifies the 
Mott-Hubbard insulator.  For the attractive interaction initially $g_s<1<g_c$, and 
both constants $U_{c,s}$ flow vice versa resulting in the spin gap insulator and
TLL of some $g_c>1$ in the charge mode. Both cases of the interaction remain 
to some extent symmetrically conjugated under the spin-charge exchange 
even after accounting for a weak backscattering
on a point impurity potential inside the wire $0<x_0<L$:
\begin{equation}
{\cal L}_{imp}= - \frac{2 V_{imp}}{\pi \alpha} 
\cos({\phi_c(t,x_c) \over \sqrt{2}} + \varphi_0 ) 
\cos({\phi_s(t,x_s) \over \sqrt{2}})
\label{2}
\end{equation}
where $x_{c,s}=x_0/v_{c,s}$, $\varphi_0\equiv \varphi + \mu x_c$ includes
a phase of the scatterer $\varphi$. The amplitude of the potential
$V_{imp}$ specifies transmittance coefficient \cite{weiss} as
$1/(1+V_{imp}^2)$, and 
$\alpha \simeq 1/E_F$ is momentum cut-off assumed to be determined by the
Fermi energy. Below  
we will first elaborate the case of the Mott-Hubbard insulator adjusting
results to the transport through the spin gap insulator at the end.

\emph{Duality Transform} - An effective model for energies lower than some
cut-off $D'$ specified below may be read off following Schmid \cite{schmid}
from the expression for the partition function ${\cal Z}$ associated to
the Lagrangian (\ref{1}) plus (\ref{2}). Without impurities the spin and charge
modes are decoupled. After integrating out $\phi_c$ in the reservoirs the charge 
mode contribution into ${\cal Z}$ describes rare tunneling between neighbor
degenerate vacua of the massive charge mode in the wire 
characterized by the quantized values of
$\sqrt{2} \phi_c(\tau,x)+ 2 \mu x = 2 \pi m $, $m$ is integer.  
Variation of $m$ by $a=\pm 1$ relates to passage of a (anti)soliton through
the wire ((anti)-instanton in imaginary time $\tau$). The tunneling amplitude
has been found \cite{1} by mapping onto a free fermionic model to be
$P e^{-s_0/T_L}, s_0=\sqrt{M^2-\mu^2} \equiv M \sin\varpi, (\mu \ll M)$
with the pre-factor $P$ and the energy cut-off $D'$ proportional to $T_L$
on approaching the perturbative regime $T_L \sim M$. The calculation \cite{2}
by instanton techniques has corrected 
$P=C \times \sqrt{D'} (\sin^3\varpi M T_L)^{1/4}$ with the constant 
$C$ of the order of 1.
The parameter $D'$ is a high-energy cut-off to the long-time asymptotics
of the kink-kink interaction: $F(\tau)= \ln\{\sqrt{\tau^2 + 1/ D'^2}\}$ created
by the reservoirs. It varies 
with $\mu$ from $D' \simeq \sqrt{M T_L}$ at $\mu=0$ 
to $ D' \simeq (M/\mu) T_L$ if $\mu > T_L$.
A crucial modification to this consideration produced by the impurity
under the assumption $E_F V_{imp} \ll M$ ensues from the shift of the 
$m$-vacuum. Since it is equal to
$(-1)^m \frac{2 V_{imp}}{\pi \alpha}  \cos \varphi  
\cos({\phi_s(\tau,x_s) \over \sqrt{2}})$ 
the neighbor vacua become non-degenerate. This 
can be accounted for by ascribing opposite
values of the pseudospin variable $\sigma = \pm 1$ to the neighbor vacua
which are the eigenvalues of the third component of the Pauli matrix $\sigma_3$.
The energy splitting becomes an operator 
$\sigma_3 \frac{2 V_{imp}}{\pi \alpha}  \cos \varphi
\cos({\phi_s(\tau,x_0) \over \sqrt{2}})$
acting on the pseudospins, and every (anti-)instanton tunneling
rotates a $\sigma_3$-value into its opposite with the Pauli matrix $\sigma_1$.
The partial function then can be written as
\endmulticols
\vspace{-6mm}\noindent\underline{\hspace{87mm}}
\begin{eqnarray}
{\cal Z} \propto \sum_{N=0}^\infty \sum_{a_j=\pm}\int D \phi_s 
{e^{-{\cal S}_0[\phi_s]} \over N!}
Tr_\sigma \left[
T \left\{\int \left(\prod_{i=1}^N d \tau_i P e^{-s_0/T_L} \sigma_1(\tau_i)\right)
\right. \right.  \nonumber \\
\times \left. \left.
\exp[ \sum_{i,j} {a_i a_j \over 2}  F(\tau_i-\tau_j) + 
\frac{2 \cos \varphi V_{imp}}{\pi \alpha}   \int d \tau \sigma_3(\tau)
\cos({\phi_s(\tau,x_0) \over \sqrt{2}})] \right\} \right]
\label{3}
\end{eqnarray}
\noindent\hspace{92mm}\underline{\hspace{87mm}}\vspace{-3mm}
\multicols{2}
\noindent
Here
${\cal S}_0[\phi_s]=\int_0^\beta d \tau \int dx \{(\partial_\tau \phi_s(\tau, x))^2
+(\partial_x \phi_s(\tau, x))^2 \}/(8\pi)$ is the free TLL Eucleadian action.
All $\tau$-integrals run from $0$ to inverse temperature
$\beta=1/T$  and $\sum_j a_j=0$. To have all 
$\sigma_{1,3}$-matrices time-ordered under the sign $T$, we attributed
each of them to a corresponding time $\tau$ assuming that 
their time evolution is trivial.  Noticing that 
the interaction $F$ coincides with the pair correlator of 
some bosonic field $\theta_c$ whose evolution is described with the
free TLL action
${\cal S}_0[\theta_c]$,
we re-write (\ref{3}) ascribing a factor
$exp(\mp \theta_c(\tau_j,0)/\sqrt{2})$
to the $\tau_j$ (anti-)instanton, respectively:
\endmulticols
\vspace{-6mm}\noindent\underline{\hspace{87mm}}
\begin{equation}
{\cal Z}\propto Tr_\sigma [ T \left\{ 
\int D \phi_s D \theta_c
exp\{- {\cal S}_0[\phi_s]-{\cal S}_0[\theta_c]
+ 2 \int d \tau [
P e^{-{s_0 \over T_L}} \sigma_1(\tau)\cos({\theta_c(\tau_j,0)\over \sqrt{2}}))
+ \frac{ \cos \varphi V_{imp}}{\pi \alpha} \sigma_3(\tau)
\cos({\phi_s(\tau,x_s) \over \sqrt{2}})]\} \right\}]
\label{4}
\end{equation}
\noindent\hspace{92mm}\underline{\hspace{87mm}}\vspace{-3mm}
\multicols{2}
\noindent
It is easy to recognize a standard Hamiltonian form 
${\cal Z}=cst \times Tr\{ e^{-\beta {\cal H}}\}$  in (\ref{4}) with
\endmulticols
\vspace{-6mm}\noindent\underline{\hspace{87mm}}
\begin{equation}
{\cal H}={\cal H}_0[\phi_s(x)]+{\cal H}_0[\theta_c(x)]
- 2 P e^{-s_0/T_L} \sigma_1 \cos(\theta_c(0)/\sqrt{2})
- \frac{2 V_{imp}}{\pi \alpha} \cos( \varphi ) \sigma_3
\cos({\phi_s(x_s) \over \sqrt{2}})
\label{5}
\end{equation}
\noindent\hspace{92mm}\underline{\hspace{87mm}}\vspace{-3mm}
\multicols{2}
\noindent
Here $\phi_s(x)$ and $\theta_c(x)$ are Schr\"{o}dinger's bosonic
operators related to the variables 
$\phi_s(\tau, x)$ and $\theta_c(\tau,x)$ of the functional integration
in (\ref{4}). The operator ${\cal H}_0[\phi_s(x)] \  
({\cal H}_0(\theta_c(x), v_c))$, a function of 
the field $\phi_s(x) \ (\theta_c(x))$ and its conjugated
is a free TLL 
Hamiltonian ($g=1$) corresponding to the free TLL action
${\cal S}_0[\phi_s]
 \ ({\cal S}_0[\theta_c])$ in (\ref{4}), respectively. 
The Dual model specified by (\ref{5}) is  equivalent to
the initial one (\ref{1}) at low energy. It relates to a
Point Scatterer with internal degree of freedom in TLL. Fortunately, this in
general rather complicated model may be solved easily through fermionization
in our particular case. This simplification stems from the marginal
behavior of the Mott-Hubbard insulator: the charge of the transport
carriers does not change on passing from the low energies to the higher ones 
despite the nature of the carriers does.

\emph{Fermionization} - From the commutation relations and hermiticity, the
Pauli matrices can be written as $\sigma_\alpha = (-1)^{\alpha+1}{i \over 2}
\sum_{\beta, \gamma} \epsilon^{\alpha, \beta, \gamma} \xi_\beta \xi_\gamma$ 
with Majorana fermions $\xi_{1,2,3}$ and antisymmetrical tensor 
$\epsilon: \ \epsilon^{123}=1 $. Since the interaction in (\ref{5}) is point-like
localized and its evolution involves only the appropriate time-dependent
correlators, we can fermionize it making use of:
\[
\psi_c(0)=\sqrt{D' \over 2\pi }\xi_1 e^{i{ \theta_c(0) \over \sqrt{2}}}, \ 
\psi_s(0)=\sqrt{E_F \over 2 \pi}\xi_3 e^{i{\phi_s(x_s) \over \sqrt{2}}}
\]
Here $\psi_{c,s}(0)$ is the $x=0$ value of the charge (spin) fermionic field, 
respectively. These fields have linear dispersions 
taken after the related bosonic fields  with momentum cut-offs (equal to the 
energy ones) $D'$ and  $E_F$, respectively. 
Substitution of these fields into (\ref{5}) 
produces a free-electron Hamiltonian 
where the interaction reduces to tunneling between the $\psi_{c,s}$ 
fermions and the Majorana one $\xi_2\ (\equiv \xi$, below). Application of a
voltage $V$ between the left and right reservoirs forces us to use the
real-time representation.  Since each instanton tunneling transfers 
charge $\Delta \phi_c/(\sqrt{2} \pi)=1e$ between the reservoirs
the voltage may be accounted for
by a shift $\theta_c/  \sqrt{2} \rightarrow \theta_c / \sqrt{2}+ Vt$  of  
the $\cos$-argument in the real-time form of the action from Eq.(\ref{4}). 
Assuming that it remains small enough, $V<T_L<M$, we will neglect its effect
on the other parameters. The real-time Lagrangian
associated with the fermionized Hamiltonian (\ref{5}) reads: 
\endmulticols
\widetext
\vspace{-6mm}\noindent\underline{\hspace{87mm}}
\begin{equation}
{\cal L}_{F}=i \xi \partial_t \xi(t) + i\sum_{a=c,s} 
\int dx \psi^+_a (\partial_t + \partial_x) \psi_a
-\sqrt{\Gamma_2 } [\psi^+_c(0,t)\xi(t) e^{-iVt}+h.c.]
-\sqrt{\Gamma_1 }[\psi^+_s(0,t) \xi(t) +h.c.]
\label{6}
\end{equation}
\noindent\hspace{92mm}\underline{\hspace{87mm}}\vspace{-3mm}
\multicols{2}
\noindent
where the rate of impurity scattering is 
$\Gamma_1=\frac{2E_F}{\pi} (\cos( \varphi ) V_{imp})^2 $
and the rate of the instanton tunneling is
$\Gamma_2=2 \pi C^2 \sqrt{T_L M \sin^3 \varpi} e^{-2s_0/T_L}$. The current 
flowing through the channel is  
$J=-\frac{\partial {\cal L}_F}{\partial(Vt)}=
-i\sqrt{\Gamma_2 }[\psi^+_c(0,t)\xi(t) e^{-iVt}-h.c.]$.
Its calculation with the non-equilibrium Lagrangian (\ref{6}) needs avail
of the Keldysh technique. Transforming voltage into 
the non-zero chemical potential of the fermions equal to $-V$, 
we can write in standard notations \cite{mah} 
\begin{eqnarray}
J=2 \sqrt{\Gamma_2} \int {d \omega \over 2 \pi} 
Re G^>_{\psi_c \xi}(\omega )
=\Gamma_2 \int {d \omega \over 2 \pi} Im \left[\left(2 \right.\right.
\nonumber \\
\times \left. \left.
 \left(1-f\left({\omega +V \over T}\right)\right)G^A_\xi(\omega)
+G^>_{\xi}(\omega )\right) \right]
\label{7}
\end{eqnarray}
where $f$ is the Fermi distribution function. To find the Green 
functions $G^A_\xi, \ G^>_{\xi}$ of the Majorana field, its free 
Green functions:$g^{R(A)}_\xi =2/(\omega \pm i0), \ 
g^{<,>}_\xi = \pm 2 \pi i \delta(\omega)$ are substituted
into the appropriate Dyson equations \cite{mah}:
\begin{eqnarray}
G^{R(A)}=g^{R(A)}(1+\Sigma^{R(A)}G^{R(A)}) \nonumber \\
G^\gamma= (1+G^R \Sigma^R )g^\gamma (1+\Sigma^{A}G^{A})+
G^R \Sigma^\gamma G^A
\label{8}
\end{eqnarray}
where $\gamma $ stands for $<,>,c,\tilde{c}$. The self-energies for the 
Majorana field are:$\Sigma^{R(A)}_\xi=\mp i(\Gamma_1+\Gamma_2), \ 
\Sigma^>_\xi=-i(\Gamma_2(2-\sum_\pm f((\omega \pm V)/T))+2(1-f(\omega/T))$.
Their substitution into (\ref{8}) and into (\ref{7}), subsequently,
results in the current:
\begin{equation}
J=\frac{2 \Gamma_2(\Gamma_1+\Gamma_2)}{\pi}\int d\omega 
\frac{f({\omega-V \over T})-f({\omega +V \over T})}{\omega^2+4(\Gamma_1+\Gamma_2)^2}
\label{9}
\end{equation}
which is the current passing through a resonant level 
of the half-width $2(\Gamma_1+\Gamma_2)$ and suppressed by the factor
$\Gamma_2/(\Gamma_1+\Gamma_2)$. The typical features of this current 
can be illustrated with its zero temperature behavior vs. voltage
and the linear bias conductance $G$ vs. temperature, respectively:
\begin{eqnarray}
J={2 \Gamma_2 \over \pi }\arctan \left(\frac{V}{2(\Gamma_1+\Gamma_2)}\right)
\label{10} \\
G= {\Gamma_2 \over \pi^2 T } 
\psi'\left({1 \over 2}+\frac{\Gamma_1+\Gamma_2}{\pi T}\right)
\label{10'}
\end{eqnarray}
where $\psi'(x)$ is the derivative of the di-gamma function,
$\psi'(1/2)=\pi^2/2$, and
the high temperature asymptotics of $G$ is $\Gamma_2 /(2T)$. 
The zero-temperature conductance 
$G \rightarrow \frac{\Gamma_2}{\pi (\Gamma_1+\Gamma_2)}$ 
manifest in (\ref{9}) is the function of $\Gamma_1/\Gamma_2 \propto
(\cos \varphi V_{imp} e^{s_0/T_L})^2 E_F/\sqrt{T_L M \sin^3 \varpi}$.
Comparison with the initial transmittance of the impurity scatterer
shows that $\sqrt{\Gamma_1/\Gamma_2}$ may be conceived as a 
renormalization of the initial amplitude $V_{imp}$ by the 
instanton exponent and a power factor of the TLL for $g=1/2$.
 
\emph{Spin gap insulator} - In this case
the chemical potential of the spin mode always
lies in the center of a gap $2M$. Meanwhile the charge mode
is inhomogeneous TLL with the constant $g_c\equiv g>1 (g_c=1)$ inside 
(outside) the wire. 
Under Duality Transform applied to the massive spin mode the partition
function ${\cal Z}' $ takes
the form of Eq.(\ref{4}) with the following
modifications. The field $\phi_s$ ($\theta_c$) is  changed by 
$\phi_c$ ($\theta_s$), respectively, and $x_s$ by $x_c$. 
The factor $\cos \varphi$ disappears,
as $\varphi$ is accumulated by $\phi_c$. The amplitude of the instanton 
tunneling between the spin mode vacua becomes $P' e^{-M/ T_L'}$, where  a new 
pre-factor $P'$ and cut-off $D'\!'$ are specified by the same expressions
which were written for $P$ and $D'$, respectively, 
after substitution of $T_L'$ instead of $T_L$ into them. 
The action for the $\phi_c$
field is inhomogeneous:
${\cal S}[\phi_c]=\int_0^\beta d \tau \int dx \{(\partial_\tau \phi_c(\tau, x))^2
+(\partial_x \phi_c(\tau, x))^2 \}/(8 g_c(x)\pi)$. To proceed in line
with the above scheme of calculation, 
we have to first integrate out the quick modes of 
the $\phi_c$ field whose energies are larger than $T_L$. Unless $v_c/v_s$ 
becomes essentially small at $g \gg 1$, these energies lie much
higher than the expected crossover. Therefore, we can 
integrate neglecting the weak tunneling between the spin mode vacua.
In the lowest order in $V_{imp}^2$ variation of the cut-off from $E_F$
to $T_L$ results in multiplication of $V_{imp}^2$
by a factor:$(E_F/T_L)^{2-g} F(x_c T_L)$ where $F(x_c T_L)$ accounts for the
interference produced by the inhomogeneity of the forward scattering.
This function may be extracted from the long-time asymptotics of 
$<e^{i \phi_c(x,t)} e^{-i \phi_c(x,0)}>  $
calculated \cite{we} with ${\cal S}[\phi_c]$ as: $F^2(z)=
const(z^2+\gamma^2)^{gr}\prod_{\pm,m=1}^\infty[(m\pm z)^2+\gamma^2]^{gr^{2m\pm 1}} 
\approx const' \times [(z^2+\gamma^2)((1-z)^2+\gamma^2)]^{gr}$,
where $\gamma=T_L/E_F$ and $r={1-g \over 1+g}$ is a negative reflection coefficient.
At the energies below the new cut-off $T_L$, the inhomogeneous action 
${\cal S}[\phi_c]$ may be changed by the homogeneous one ${\cal S}_0[\phi_c]$.
Then exercising the above fermionization we come to a real-time Lagrangian 
analogous with (\ref{6}):
\endmulticols
\widetext
\vspace{-6mm}\noindent\underline{\hspace{87mm}}
\begin{equation}
{\cal L}_{F}=i \xi \partial_t \xi(t) + i\sum_{a=c,s} 
\int dx \psi^+_a (\partial_t +  \partial_x) \psi_a
-\sqrt{\Gamma'_1 } [\psi^+_c(0,t)\xi(t) e^{-iVt}+h.c.]
-\sqrt{\Gamma'_2 }[\psi^+_s(0,t) \xi(t) +h.c.]
\label{11}
\end{equation}
\noindent\hspace{92mm}\underline{\hspace{87mm}}\vspace{-3mm}
\multicols{2}
\noindent
Here the rate of the instanton tunneling 
$\Gamma'_2=2 \pi C^2 \sqrt{T'_L M } e^{-2M/ T_L'}$
has an manifest similarity with $\Gamma_2$ and
the rate of impurity scattering
$\Gamma'_1=\frac{2E_F}{\pi} (E_F/T_L)^{1-g} F(x_c T_L) V_{imp}^2 $
acquires an additional small factor $(E_F/T_L)^{1-g} F(x_c T_L)$, 
as $g_c>1$. Creation and annihilation
of the fermion of the charge mode in the fermionized model describes 
a change of chirality of one quasiparticle inside the wire. 
Therefore, the term proportional to $\sqrt{\Gamma'_1 }$
in (\ref{11}) leads to the  backscattering current:
$J_{bsc}=-\frac{\partial {\cal L}_F}{\partial(Vt)}=
-i\sqrt{\Gamma'_1}[\psi^+_c(0,t)\xi(t) e^{-iVt}-h.c.]$ equal to minus deviation
of the direct current from its maximum $V/\pi$ value.
The new Lagrangian (\ref{11}) and the current $J_{bsc}$ 
transform into the old Lagrangian (\ref{6}) and the current $J$,
respectively, on changing $\Gamma_1', \Gamma_2'$ by $\Gamma_2, \Gamma_1$.
Hence the average value of $J_{bsc}$ may be extracted from (\ref{9})
in the same way. In particular, the linear bias conductance:
$G=(1- {\Gamma'_1 \over  \pi T } 
\psi'\left(1 / 2+\{\Gamma_1'+\Gamma_2'\}/(\pi T)\right))/\pi$ is strongly
suppressed at $T=0$ as $G=\frac{\Gamma_2'}{\pi (\Gamma_1'+\Gamma_2')}$. 
Similarly to the Mott-Hubbard case it shows that the 
initial amplitude of the potential $V_{imp}$ renormalizes to $\sqrt{\Gamma_1'/\Gamma_2'} \propto
V_{imp} e^{M/ T_L'} \sqrt{E_F (E_F/T_L)^{1-g_c} F(x_c)/\sqrt{T'_L M}}$ at low energy. Above the
crossover temperature $2(\Gamma'_1+\Gamma'_2)$ the conductance recovers
as ${1 \over \pi}\left(1-{\Gamma'_1 \over 2T} \right)$.
The backscattering current related to (\ref{10})
results in the constant shift of the direct current $J=V/\pi -\Gamma'_1$
above the voltage crossover $2(\Gamma'_1+\Gamma'_2)$.

Finally, we have shown that a zero-energy resonance in transport
through a 1D Mott-Hubbard insulator of finite length 
predicted \cite{1} for the clean wire stands weak impurity
pinning while the rate of the impurity backscattering $\Gamma_1$ is small
$\Gamma_1 \ll T_L$. We have also described how opening a gap in the spin mode 
suppresses the charge transport.

This work was supported by the Center of Excellence.

$^*$ On leave of absence from A.F.Ioffe Physical Technical 
Institute, 194021, St. Petersburg, Russia.\\
Present address: Departmetn of Physics and Astronomy, SUNY at Stony Brook,
NY 11794, USA.


\begin{references}

\bibitem{1}V. V. Ponomarenko and N. Nagaosa, Phys. Rev. Lett. 
{\bf 81}, 2304 (1998).

\bibitem{hald}F.D.M. Haldane, Phys. Rev. Lett.{\bf 47}, 1840 (1981).

\bibitem{tar}S. Tarucha, T. Honda, and T. Saku,  Solid State Commun.
{\bf 94}, 413 (1995).

\bibitem{ya} A. Yacobi {\it et al}, Solid State Commun.
{\bf 101}, 77 (1996); Phys. Rev. Lett. {\bf 77}, 4612 (1996).

\bibitem{ftr} H. Fukuyama, S.Tarucha (private communications);
A.A.Odintsov, Y.Tokura and S.Tarucha, 
Phys. Rev. B {\bf 56}, R12729 (1997).


\bibitem{sol} Solyom, Adv. Phys. 31, 293 (1979).

\bibitem{2}V. V. Ponomarenko and N. Nagaosa, cond-mat/9806136;
Phys. Rev. Lett. {\bf 83}, 1822 (1999).

\bibitem{schmid}A. Schmid, Phys. Rev. Lett. {\bf 79}, 1714 (1984).

\bibitem{jer}D. Jerom and H. Schults, Adv. Phys. 31, 293 (1982).

\bibitem{weiss} U. Weiss, Solid State Commun. {\bf 100}, 281 (1996).

\bibitem{mah}G.D. Mahan, {\ Many-Particle Physics }, 
(Plenum Press, New York, 1990). 

\bibitem{we}V. V. Ponomarenko and N. Nagaosa, Phys. Rev. Lett. 
{\bf 79}, 1714 (1997).

\end{references}
\end{document}